# Charge-density-wave resistive switching and voltage oscillations in ternary chalcogenide BaTiS$_3$

Huandong Chen[1], Nan Wang[1], Heifei Liu[2], Han Wang[1,2], Jayakanth Ravichandran[1,2,3*]

[1]Mork Family Department of Chemical Engineering and Materials Science, University of Southern California, Los Angeles, California, USA

[2]Ming Hsieh Department of Electrical and Computer Engineering, University of Southern California, Los Angeles, California, USA

[3]Core Center of Excellence in Nano Imaging, University of Southern California, Los Angeles, California, USA

*e-mail: j.ravichandran@usc.edu

**Abstract**

Phase change materials, which show different electrical characteristics across the phase transitions, have attracted considerable research attention for their potential electronic device applications. Materials with metal-to-insulator or charge density wave (CDW) transitions such as $VO_2$ and 1$T$-$TaS_2$ have demonstrated voltage oscillations due to their robust bi-state resistive switching behavior with some basic neuronal characteristics. $BaTiS_3$ is a small bandgap ternary chalcogenide that has recently reported the emergence CDW order below 245 K. Here, we report on the discovery of DC voltage / current-induced reversible threshold switching in $BaTiS_3$ devices between a CDW phase and a room temperature semiconducting phase. The resistive switching behavior is consistent with a Joule heating scheme and sustained voltage oscillations with a frequency of up to 1 kHz has been demonstrated by leveraging the CDW phase transition and the associated negative differential resistance. Strategies of reducing channel sizes and improving thermal management may further improve the device performance. Our findings establish $BaTiS_3$ as a promising CDW material for future energy-efficient electronics, especially for neuromorphic computing.

**Main Content**

The metal-to-insulator transition is a hallmark phenomenon predicted by Peierls' theory for explaining charge density wave (CDW) in ideal one-dimensional metals[1,2]. However, this transition is not a universal feature in real CDW materials, with some systems failing to exhibit transport anomalies at the transition temperatures[3,4]. Despite over half a century of study into CDW phases and phase transitions, much of the research has centered on the physical aspects such as the driving force behind CDW[5-7] and its relation to superconductivity[8-10], rather than exploring potential electronic device applications. Nonetheless, CDW systems with hysteric resistive phase transitions have the potential to offer unique opportunities for novel electronic device development[11,12]. One such system that has attracted significant attention is the quasi-two-dimensional Mott insulator 1$T$-TaS$_2$, which exhibits several CDW phase transitions with resistivity changes and hysteresis and has been utilized to construct electronic devices such as phase change oscillators[13-15] and memristors[16,17].

BaTiS$_3$ is a quasi-one-dimensional small bandgap semiconductor[18] that has recently shown unique resistive phase transitions[19], including one CDW transition near 250 K and another structural transition emerging at even lower temperatures (from 120 K to 150 K during cooling cycle). Upon cooling from room temperature, the system switches from a semiconducting state to a CDW state, resulting in an increase in electrical resistivity, bandgap opening, and a periodic lattice distortion[19]. This change in electrical resistivity presents opportunities to create devices that can be modulated by external stimuli such as electrical and optical fields. In this study, we demonstrate threshold resistive switching behavior with negative differential resistance (NDR) utilizing the CDW phase transition in bulk BaTiS$_3$ single crystal. The electrical switching mechanism was extensively investigated through temperature-dependent current-voltage (I-V) characteristics and pulsed I-V measurements. Furthermore, voltage oscillations with a frequency close to 1 kHz were observed from a two-terminal BaTiS$_3$ device. Potential strategies for optimizing device performance, such as reducing channel sizes and optimizing thermal management, were also explored. Our findings shed light on electronic device applications in CDW systems.

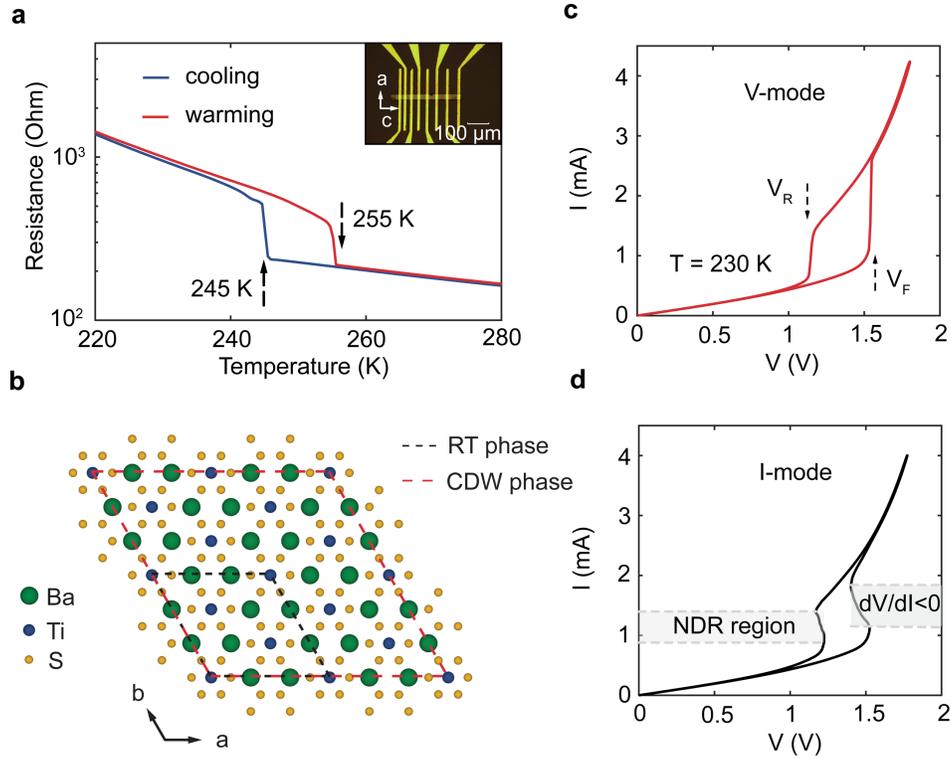

**Fig. 1 Transport anomalies and reversible resistive switching in BaTiS$_3$** (**a**) Representative temperature dependent electrical resistance of BaTiS$_3$ crystal along *c*-axis. Transport anomalies with abrupt and hysteric jumps near 240 -260 K reveal the existence of a phase transition. The inset shows an optical microscopic image of a typical BaTiS$_3$ device. (**b**) Illustration of unit cell evolution of BaTiS$_3$ across the CDW phase transition. (**c**) and (**d**) I-V characteristics of a two-terminal BaTiS$_3$ device at 230 K by sweep voltage and current, respectively. Negative differential resistance (NDR) regions are observed in I-mode.

## Reversible resistive switching

The phase transition of BaTiS$_3$ near 250 K, referred to as the CDW transition, results in an abrupt increase in resistance and a thermal hysteresis of over 10 K, both of which are crucial for its potential applications as an electronic device. Figure 1a plots the temperature-dependent resistance of BaTiS$_3$ along the *c*-axis from 220 to 280 K, with an inset showing an optical image of a typical multi-terminal BaTiS$_3$ device with varying channel sizes. Structurally, the unit cell doubles along the *a*- and *b*-axis across the transition ($a = 2a_0$, $b = 2b_0$, $c = c_0$), as depicted in Figure 1b.

To demonstrate the resistive switching behavior in BaTiS$_3$, the device was initially set to the high-resistive CDW state near the completion of the CDW transition (230 K). A DC current-voltage characterization was performed on a two-terminal BaTiS$_3$ device by sweeping voltage (V-mode), as illustrated in Figure 1c. The system exhibited a transition to a more conductive state above a certain threshold voltage $V_F$ during forward scan, and it returned to its original high-resistive state below another critical voltage $V_R$ ($V_R < V_F$) during reverse scan, forming a characteristic hysteresis window. Additionally, 'S-type' negative differential resistance regions ($dV / dI < 0$) were observed during transitions when testing in current mode (I-mode) by sourcing current, as shown in Figure 1d. This threshold resistive switching behavior with NDR has been previously observed in other phase change systems such as VO$_2$[20-22] and 1T-TaS$_2$[13,23,24], and is utilized to construct electronic devices such as oscillators[13,14,21,22].

## Mechanism of electrical switching

The mechanism behind such electrical voltage / current resistive switching behavior in those systems is believed to be primarily due to local Joule heating[24-26], although there are ongoing debates regarding the potential role of electrical field effect[23] and Mott transition[21]. The effect of

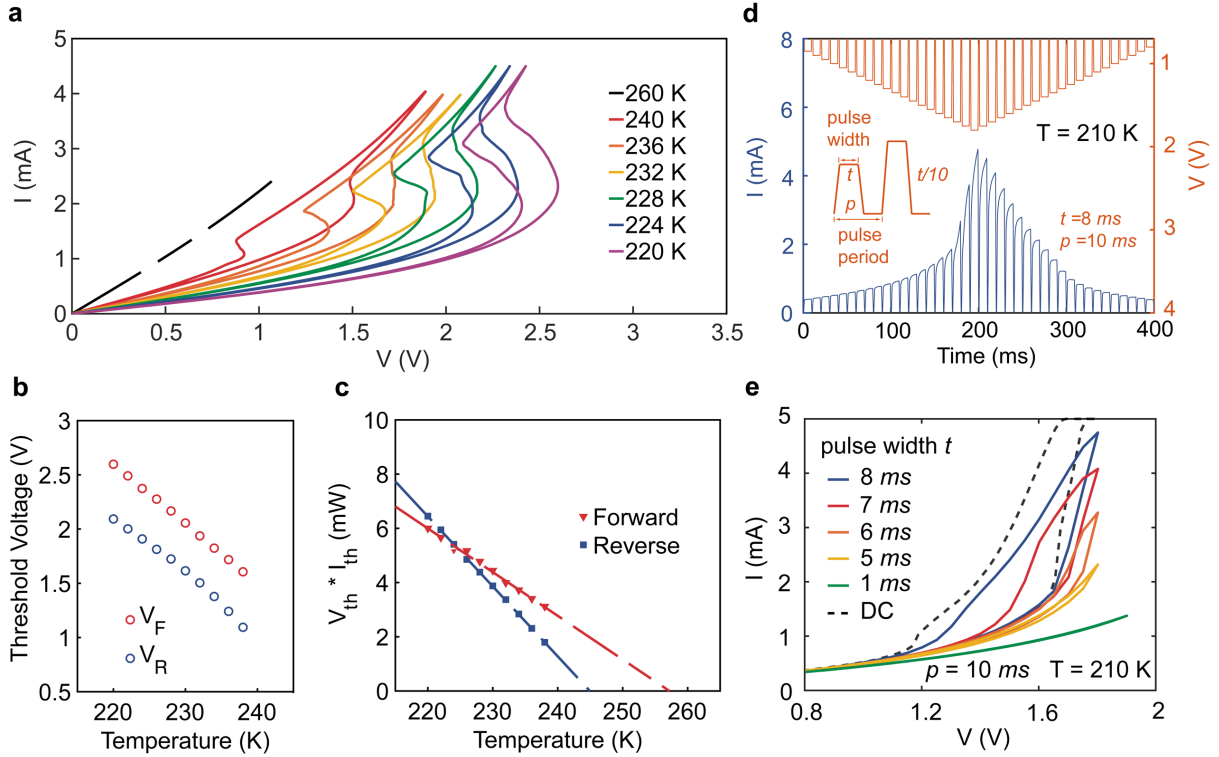

**Fig. 2 Joule heating mechanism behind CDW resistive switching in BaTiS$_3$.** (**a**) Four-probe I-V characteristics of a BaTiS$_3$ device at different temperatures. (**b**) Extracted threshold voltages at the corresponding temperatures for both forward and reverse sweeps. (**c**) Calculated temperature-dependent thermal power at threshold fields ($P_{th}$). A linear relationship is found between $P_{th}$ and temperature. (**d**) Pulsed I-V measurement of a two-terminal BaTiS$_3$ device at 210 K. The pulse voltage was ramped linearly from 0.8 V to 1.8 V, with the pulse width $t$ = 8 ms and pulse period $p$ = 10 ms. (**e**) Reconstructed I-V characteristics of BaTiS$_3$ from pulse measurements with pulse width varying from 8 ms to 1 ms, while the pulse period was maintained at 10 ms. Dashed line shows the DC I-V scan on the same device for comparison.

Joule heating is highly dependent on the specific material, device structure, and bias mode (DC or pulse). In this experiment scheme, where a BaTiS$_3$ crystal is embedded in a polymer medium with low thermal conductivity and the switching is triggered by DC sweeps, the local Joule heating can be substantial.

To evaluate the contribution of Joule heating to the resistive switching behavior observed in BaTiS$_3$, we conducted four-probe I-V sweeps at various temperatures across the CDW transition (Figure 2a). The results show that the critical voltage required to switch the resistance state increases as temperature decreases, and there is no threshold voltage switching observed at a temperature of 260 K. The thermal power generated by Joule heating at threshold fields were calculated ($P_{th} = V_{th} \times I_{th}$) and found to exhibit a linear relationship with temperature, as illustrated in Figure 2b and 2c. Two characteristic temperatures of 245 K and 258 K were identified at which the threshold thermal power approaches zero, which aligns with the transition temperatures observed in the temperature-dependent resistance measurements (Figure 1a). This analysis suggests that the resistive switching in BaTiS$_3$ is primarily driven by Joule heating.

Moreover, pulsed I-V measurements with varying pulse widths were performed to gain insights into the switching mechanism by deconvoluting the contributions from Joule heating and electrical field effects. Figure 2d shows the results of pulsed I-V measurements conducted on a two-terminal BaTiS$_3$ device at 210 K. Voltage pulses with a width of 8 ms and a pulse period of 10 ms were swept between 0.8 V to 1.8 V. In such pre-defined voltage pulses, less than 10% of the total time was used for cooling. Hence, similar to that observed in DC sweeps, the hysteretic switching behavior persisted, as evidenced by the asymmetric measured current profile. To reduce the contribution of Joule heating, the pulse width was decreased from 8 ms to 1 ms while maintaining the voltage sweep ranges and pulse period. Figure 2e illustrates the reconstructed I-V curves from pulsed measurements for different pulse widths. As the pulse width was decreased, the width of the hysteresis window reduced while the switching voltage increased. No hysteresis was revealed in the I-V curves when the pulse width was 1 ms. These observations are consistent with the hypothesis of a thermally driven transition, as the Joule heating power decreased with decreasing pulse width, while the electric field applied to the BaTiS$_3$ device remained the same.

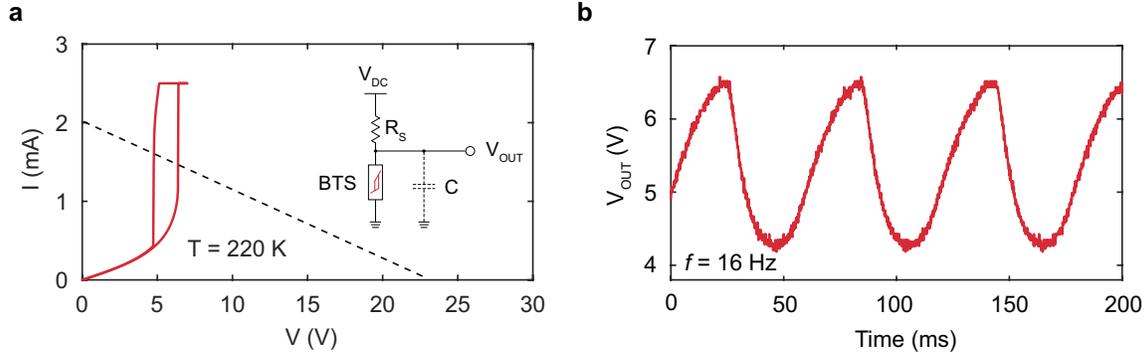

**Fig. 3 CDW voltage oscillation in bulk BaTiS$_3$**. (**a**) I-V characteristics of a two-terminal BaTiS$_3$ device at 220 K, with the inset showing the circuit for oscillation measurements. (**b**) Representative oscillation waveform of BaTiS$_3$ with a frequency of 16 Hz.

## CDW voltage oscillations

The capability of inducing sustained phase change-based voltage oscillations is crucial for constructing electronic devices such as oscillators. Systems such as 1$T$-TaS$_2$ and VO$_2$ have shown promising results, with the potential for GHz-level switching speeds theoretically predicted[25,27] and experimentally demonstrated at MHz frequencies[13,21]. In the case of BaTiS$_3$, we observe stable voltage oscillations at frequencies from 16 Hz to 18 Hz when a two-terminal BaTiS$_3$ device is connected in series with a load resistor ($R_L$ = 11.25 kOhm) and a parallel capacitor ($C_P$ = 10 μF) and subjected to a DC bias between 16 V and 23 V. The voltage oscillation is illustrated in Figure 3b, with the I-V characteristics at 220 K and circuit diagram of the oscillation measurements shown in Figure 3a, together with the load line of the resistor. The mechanism behind this voltage oscillation can be understood as the switching of the BaTiS$_3$ device between its CDW state and semiconducting state. When the DC voltage reaches a critical value $V_F$, the voltage drop across the BaTiS$_3$ channel triggers a transition to the semiconducting state, resulting in a sudden increase in current and a subsequent increase in voltage across the load resistor. This drives the BaTiS$_3$ device back into the CDW state, and the cycle repeats, leading to sustained voltage oscillations. The oscillation is not sustained when the DC voltage exceeds 23 V or falls below 16 V.

On the other hand, although stable voltage oscillations were obtained for the first time from the CDW transition in BaTiS$_3$, its frequency remains orders of magnitude lower than the MHz level achieved in VO$_2$ and 1$T$-TaS$_2$ systems. This low switching speed is primarily attributed to the poor heat dissipation of the bulk BaTiS$_3$-polyimide system, which has a low thermal conductivity and results in an inefficient cooling process that limits its overall performance, despite the effective heating procedure.

## Effect of thermal management and channel sizes

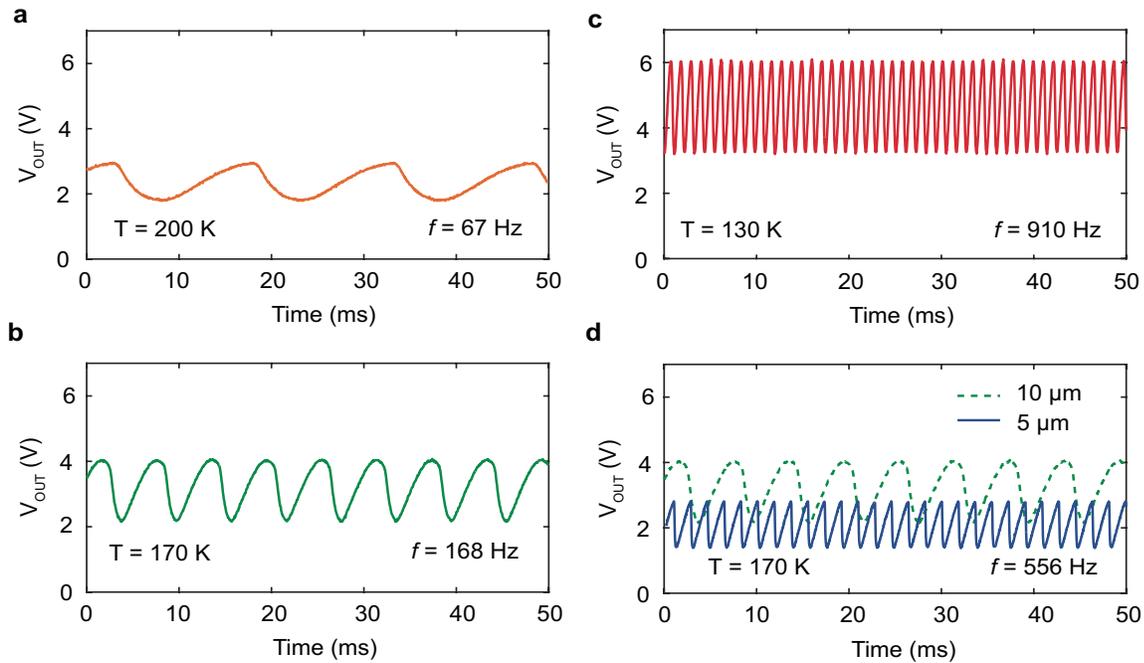

**Fig. 4 Oscillation frequency optimization in BaTiS$_3$.** (**a**) to (**c**) Effect of operating temperature. The frequency increases from 67 Hz to 910 Hz by reducing the measurement temperature from 200 K to 130 K. (**d**) Effect of device channel size. The oscillation frequency increases more than three times when reducing the channel size from 10 μm to 5 μm.

One approach to enhance the switching speed is by improving the cooling efficiency of the system, for example, by decreasing the operating temperature. Figure 4a to 4c plot the voltage oscillations of the same $BaTiS_3$ device measured at 200 K, 170 K and 130 K, respectively, with an observed increase in frequency from 67 Hz to 910 Hz. However, it is difficult to maintain this CDW oscillation when the temperature is further reduced, as the low-temperature structural transition in $BaTiS_3$ begins to interfere and complicate the results. Thus, it is challenging to significantly improve the oscillator performance solely by tuning the measurement temperatures.

Another strategy to further enhance the oscillation frequency is to decrease the size of the device channel, as has been proven effective in other oscillating systems, such as $VO_2$[28]. In early days, the voltage oscillation frequency in millimeter-scale $VO_2$ bulk single crystals was only around 5 kHz[29], while nowadays, MHz-level oscillation frequencies have been achieved in $VO_2$ thin film devices with sub-micron channels[30]. Figure 4d plots the oscillation waveforms from two different $BaTiS_3$ devices with channel sizes of 10 μm and 5 μm, respectively, both of which were measured at 170 K for direct comparison. With reduced channel size down to 5 μm, the oscillation frequency increases more than three times compared to the 10 μm $BaTiS_3$ device. This effect can be understood as the improved efficiency for both cooling and heating processes as the channel size decreases. Further reduction of sample sizes both laterally and vertically is expected to result in even higher oscillation frequencies in $BaTiS_3$.

## Conclusion

In conclusion, we have shown reversible threshold switching in the recently discovered CDW system $BaTiS_3$, driven by DC voltage or current, whose mechanism is consistent with a Joule heating scheme. Moreover, sustained voltage oscillations were achieved in $BaTiS_3$ based on bistate resistive switching between the semiconducting phase and CDW phase. The oscillation frequencies were improved through appropriate thermal managements and reduced channel sizes. Our work on BaTiS$_3$ opens new opportunities in electronic device applications of CDW phase change materials beyond 1$T$-TaS$_2$.

## Methods:

### Device fabrication

Single crystals of BaTiS$_3$ were grown by a vapor transport method as reported elsewhere[18,19]. Needle-like crystals with thickness of 5-15 µm were individually picked and embedded in low-stress polymer medium for planarization[19,31]. Multi-terminal BaTiS$_3$ devices were fabricated using standard photolithography and ebeam evaporation (Ti / Au = 3 / 300 nm). An SF$_6$ / Ar reactive ion etching (RIE) treatment (SF$_6$ / Ar = 15 / 50 sccm, 100 W, 100 mTorr, 1 min) was applied right before metallization to remove surface oxides and reduce the contact resistance.

### Electrical characterization

Electrical transport measurements were performed in a JANIS 10 K closed-cycle cryostat. Temperature dependent resistances of BaTiS$_3$ as shown in Figure 2a were measured using standard low-frequency ($f$ = 17 Hz) AC lock-in techniques (Stanford Research SR830) in four-probe geometry, with an excitation current of about 1 µA. All DC current-voltage (I-V) characteristics in sweep mode were measured with a semiconductor analyzer (Agilent 4156C). The oscillation measurements were conducted using a source meter (Keithley 2400) to supply constant voltage /

current and a digital oscilloscope (Keysight DSOX1204G) to record the waveforms of voltage oscillations. Pulsed I-V characteristics were measured using a semiconductor analyzer (Keysight B1500A) equipped with a WGFMU (waveform generator / fast measurement unit) module.

## Author declarations

### Conflict of interest

The authors declare no competing financial interests.

### Acknowledgements

We gratefully acknowledge support from an ARO MURI program (W911NF-21-1-0327), an ARO grant (W911NF-19-1-0137) and National Science Foundation (DMR-2122071).

### Author contributions

H.C. and J.R. conceived the idea and designed the experiments. H.C. fabricated devices and performed electrical characterizations, with input from N.W. H.L. helped on the pulsed I-V measurements. H.C. and J.R. wrote the manuscript with input from all other authors.

31 Chen, H., Avishai, A., Surendran, M. & Ravichandran, J. A polymeric planarization strategy for versatile multiterminal electrical transport studies on small, bulk crystals. *ACS Applied Electronic Materials* **4**, 5550-5557 (2022).
16